\pdfoutput=1
\documentclass[12pt]{iopart}
\usepackage{iopams}
\usepackage{graphicx}
\usepackage{epsfig}

\begin{document}

\title[]{Analytical solution
for the diffraction of an electromagnetic wave by a graphene grating}

\author{T.M.~Slipchenko$^1$, M.L.~Nesterov$^1$, L.~Martin-Moreno$^1$, A.Yu.~Nikitin$^{2,3}$}

\address{$^1$ Instituto de Ciencia de Materiales de Arag$\mathrm{\acute{o}}$n and Departamento de F$\mathrm{\acute{i}}$sica de la Materia Condensada,
CSIC-Universidad de Zaragoza, E-50009 Zaragoza, Spain\\
$^2$ CIC nanoGUNE Consolider, 20018 Donostia-San Sebasti$\mathrm{\acute{a}}$n, Spain\\
$^3$ Ikerbasque, Basque Foundation for Science, 48011 Bilbao, Spain}
\ead{alexeynik@rambler.ru}

\begin{abstract}
An analytical method for diffraction of a plane electromagnetic wave at periodically-modulated graphene sheet is presented. Both interface corrugation and periodical change in the optical conductivity are considered. Explicit expressions for reflection, transmission, absorption and transformation coefficients in arbitrary diffraction orders are presented. The dispersion relation and decay rates for graphene plasmons of the grating are found. Simple analytical expressions for the value of the band gap  in the vicinity of the first Brillouin zone edge is derived.
The optimal amplitude and wavelength, guaranteeing the best matching of the incident light with graphene plasmons are found for the conductivity grating. The analytical results are in a good agreement with first-principle numeric simulations.
\end{abstract}

\pacs{78.67Wj, 73,20Mf, 42.25Bf}

\section{Introduction}

Since the pioneering predictions, plasmons in graphene (GPs)~\cite{Ken, Vaf, hanson}, have been intensively studied theoretically,
~\cite{Jab, nikgr, Kop, Nik1, Nik1a, Nik1aa, bludrev, blud1, condmod, blud2}, and recently have been observed experimentally,
~\cite{Ju, hillebrandt, basov, Fei, yan}. The excitation of GPs by external radiation presents interest for both fundamental and technological aspects. The following configurations for coupling of electromagnetic radiation to GPs have been considered: graphene sheet having either modulated optical conductivity~\cite{bludrev, blud1, condmod, blud2} or relief corrugations~\cite{bludrev}; graphene monolayer placed on subwavelength dielectric gratings \cite{zhen};  patterned graphene structures, including
one-dimensional micro-ribbons \cite{Ju, bludrev, Nik2} and two-dimensional microdisk arrays, \cite{Yan, kop2}.
The majority of studies focused on numeric calculations of absorption, transmission and reflection properties.

In this paper we perform a completely analytical analysis of the vectorial diffraction problem based on the resonance perturbation theory~\cite{nik, nik3}. This method allows us to derive the transmission, reflection and absorption coefficients in a simple closed form.  Then, we find the optimal depth of the grating as a function of the wavelength, for which the maximal value of the excited GP mode is achieved. We also obtain the combination of the grating depth and wavelength which provides the absolute maximum of the GP amplitude. From the experimental point of view, this optimal combination of the parameters should lead to observation of absorption maximum and the minimum of transmission. Within the same method, the homogeneous problem for the eigenmodes of the structure is considered. We derive the grating-induced correction for the GP decay rate and the value of band gap  of the GPs at the first Brillouin zone edge.

On each stage the validity of the analytical results has been confirmed by finite-elements simulations. These simulations have been also used to extend our analytical study of graphene corrugation gratings to corrugation depths beyond the range of the validity of the perturbation theory.

\section{Description of the system and method}\label{2}

\begin{figure}
\begin{center}
\begin{center}
\includegraphics[width=8.5cm]{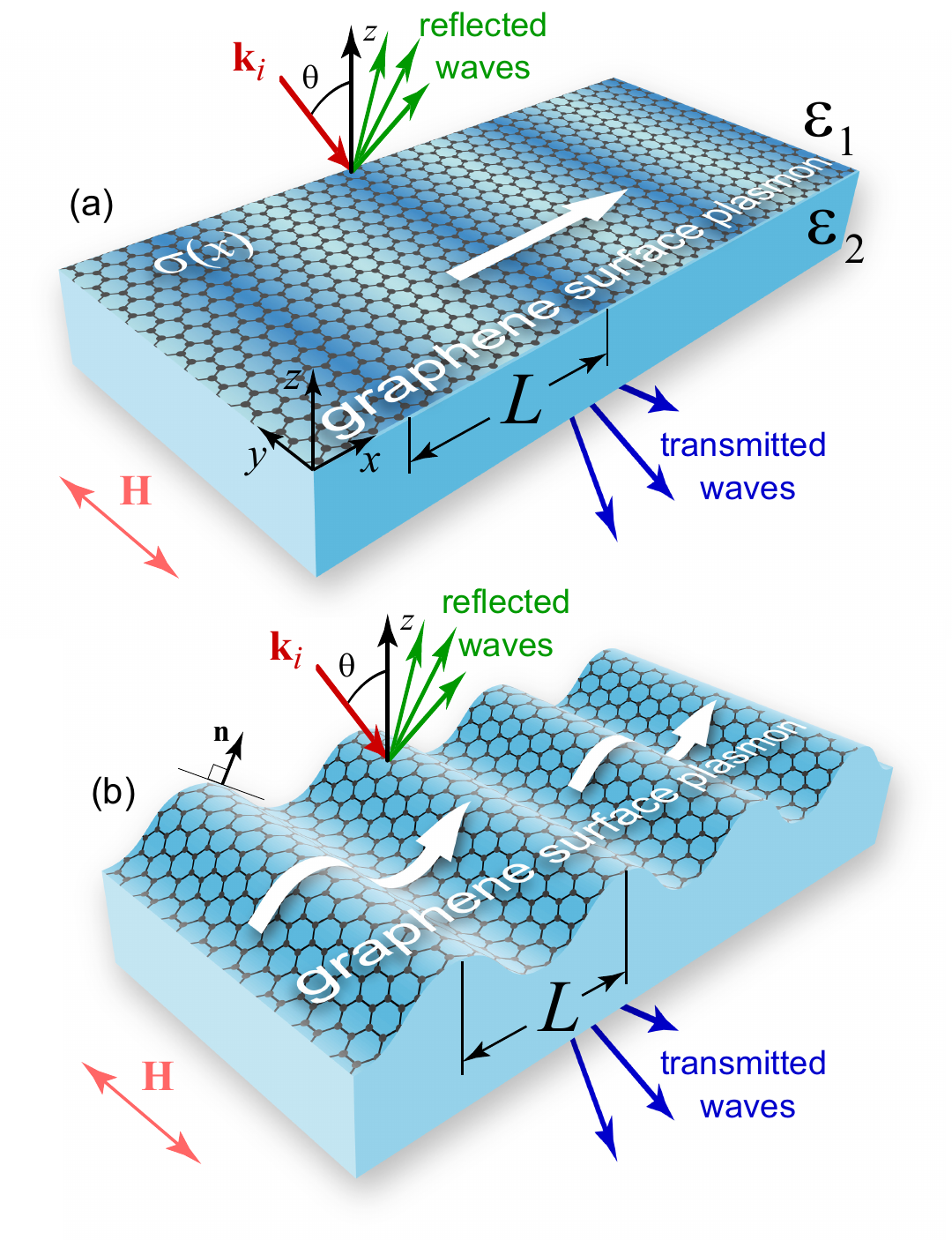}
\end{center}
\caption{(Color online) Geometry of the problem. Diffraction of a plane wave at (a) conductivity grating, (b) relief grating. }\label{graph1}
\end{center}
\end{figure}

We consider a periodic grating, formed by the variation of either conductivity $\sigma$ of graphene monolayer \cite{condmod} (see Fig.~\ref{graph1}~(a)) or the interface relief $z=\zeta(x)$ \cite{textripl} (see Fig.~\ref{graph1}~(b)) in the $x$-direction. The spacial period is $L$. The graphene sheet is placed onto the interface between two dielectrics with the permittivities $\varepsilon^{(-)}$ and $\varepsilon^{(+)}$.
In general, a graphene grating can be a combination of both the conductivity and interface variations, but in this paper we study these two cases separately. In other words, we assume that the conductivity does not depend on $x$ for the relief grating while for the conductivity grating the graphene sheet is flat.

In case of the conductivity grating, the normalized conductivity $\alpha(x)=2\pi\sigma(x)/c$  can be represented in the form of the Fourier series
\begin{equation}\label{eq1}
\alpha (x) = \alpha (x+L) = \sum_n \alpha_n e^{inGx}, \; G=\frac{2\pi}{L},
\end{equation}
The zero harmonic $\alpha_0=2\pi\sigma_0/c$ is proportional to the average conductivity $\sigma_0$. Throughout the paper the RPA conductivity model is used \cite{wunsch, hwang, falk}.
The conductivity modulation can be achieved for instance by placing the graphene into a periodic electrostatic field so that the doping level will be spatially varied \cite{condmod}. In the near infrared and THz frequencies the conductivity is
dominated by the intraband contribution and therefore it is
proportional to the Fermi energy. Then the model (\ref{eq1}) is fully justified.

When the graphene interface is periodically corrugated we have
\begin{equation}\label{eq2}
z=\zeta(x) = \zeta(x+L) =\sum_n \zeta_n e^{inGx}, \, \zeta_0=0.
\end{equation}
Notice that in the infrared and tera-hertz regimes the momentum of graphene plasmons is much larger than that of the incident plane wave. Consequently,
to compensate the momentum mismatch, the period of grating $L$ must be much smaller than
the incident wavelength, $L\ll\lambda$ (or $G\gg g$, where $g=2\pi/\lambda$).

Let a $p$-polarized plane monochromatic electromagnetic wave, with electric field $\mathbf{E}=\left\{E_x, 0, E_z\right\}$, and magnetic field
$\mathbf{H}=\left\{0, H, 0\right\}$, fields impinges onto our periodic system at an angle $\theta$ (between its wavevector and $z$-axis) from the half-space $z>0$.
The in-plane periodic modulation results in generating the infinite set of plane
diffracted waves. For the modulation of the conductivity, the grating is flat and the exact representation for the fields is given by the
Fourier-Floquet expansion. In the case of the interface corrugation we will assume that both the variation of the surface relief and its derivative
are small ($g|\zeta_n|\ll 1$ and $|\partial_x \zeta|\ll 1$). Then according to the Rayleigh approximation \cite{tzangkong} we can also use the same representation of the fields.
Thus, for both types of gratings we will take  the total magnetic field in the superstrate and substrate (referred to using the symbols "+" and "-" respectively) in the following form
 \begin{eqnarray}\label{field1}
   &H^{+}(x,z)= -\displaystyle\frac{\varepsilon^{(+)}g}{k_{z}}\exp\left[i(kx - k_{z}z)\right]+\nonumber\\
   &+\sum_n \displaystyle\frac{\varepsilon^{(+)}g}{k_{zn}^{(+)}} r_{n}^{+} \exp\left[i(k_nx+k_{zn}^{(+)}z)\right],\\
   &H^{-}(x,z)=-\sum_n \displaystyle\frac{\varepsilon^{(-)}g}{k_{zn}^{(-)}} r_{n}^{-} \exp\left[i(k_nx-k_{zn}^{(-)}z)\right],\nonumber
\end{eqnarray}
where $r_{n}^+$ and $r_n^-$ are the transformation coefficients in the superstate and substrate respectively, and the monochromatic time dependence $\exp(-i\omega t)$ is omitted.
The tangential and normal components of the wavevectors read
\begin{eqnarray*}
&k=g \sin\theta, \; k_{z}=\sqrt{\varepsilon^{(+)}g^2-k^2},\\
& k_n=k+n G, \;    k_{zn}^{(\pm)}=\sqrt{\varepsilon^{(\pm)}g^2-k_n^2},\\
& n=0, \pm1, \pm2, \ldots.
\end{eqnarray*}
The branch of square root should be chosen as $\mathrm{Im}(k_{zn}^{\pm})\geq0$, in order to satisfy the radiation conditions.
The electric field components can be readily obtained from Eq.~(\ref{field1}) using Maxwell equations $gE_x=-i\partial_z H$, $gE_z=i\partial_x H$.

The field in the upper and lower half-spaces are connected through the boundary conditions at the interface containing graphene
\begin{eqnarray}\label{bound1}
&\mathbf{E}_{t}^{+}-\mathbf{E}_{t}^{-} = 0,\\
&\mathbf{n}\times(\mathbf{H}^{+}-\mathbf{H}^{-}) = 2\alpha\mathbf{E}_t^{-}.\nonumber
\end{eqnarray}
Here $\alpha$ is defined by Eq.~(\ref{eq1}) for the conductivity grating, whereas for the case of the corrugated interface it is constant, $\alpha=\alpha_0$. The subscript $t$ stays for the tangential components of the fields and
$\mathbf{n}$ is the unitary vector normal to the surface. For the conductivity grating $\mathbf{n}=(0, 0, 1)$, while for the relief grating $\mathbf{n}=(n_x, 0, n_z)$.

The tangential component of the electric field is defined as
$\mathbf{E}_{t}=\mathbf{E}-\mathbf{E}_{n}=\mathbf{E}-\mathbf{n}\left(\mathbf{E}\cdot\mathbf{n}\right)$, where
$\mathbf{E}_{n}$ is its normal component. Projecting it onto the $x$ axis we have
${E}_{tx}=E_{x}-n_x(x)\left[n_x(x)E_{x}+n_z(x)E_{z}\right]$.
Assuming the interface corrugation to be smooth, i.e. $|\partial_x\zeta|\ll 1$, the surface normal vector can be simplified
\begin{equation*}
\mathbf{n}=\mathbf{e}_z-\mathbf{e}_x \partial_{x}\zeta + O[(\partial_{x}\zeta)^2],
\end{equation*}
so that the second term is proportional to the derivative of a small interface inclination.
We will also assume that the corrugation is shallow enough to fulfill $g|\zeta_n|\ll 1$. This allows us to greatly simplify the expressions for the fields at the interface. Indeed, in this case
the exponentials from (\ref{field1}) can be expanded into the Tailor series, and then the linear approximation is usually enough to provide precise results for shallow gratings.  The appearing exponents read
\begin{eqnarray*}
   &\exp[\pm ik_{zn}^{(\pm)}\zeta(x)]\simeq 1 \pm i k_{zn}^{(\pm)}\zeta(x),\\
  &\exp[-ik_{z}\zeta(x)]\simeq 1-i k_{z}^{(\pm)}\zeta(x).
\end{eqnarray*}
After some straightforward algebra we finally obtain an infinite liner system of equations for
plane waves amplitudes (in case of the relief grating we retain only the linear in $|\partial_x\zeta|$ and $|\zeta_n|$ terms):
\begin{eqnarray}\label{systF}
    &\sum_{m,\nu'=\pm}D_{nm}^{\nu\nu'}r_m^{\nu'}=V_n^{\nu}, \; \nu=\pm,
\end{eqnarray}
where the subscript $\nu$ denotes the field in the superstate $\nu=+$ or in the substrate $\nu=-$.
The matrix elements may be represented as the sum of the
diagonal, $b_n^{\nu\nu'}$, and off-diagonal elements, $d_{n,m}^{\nu\nu'}$ (see Appendix for more details):
\begin{eqnarray*}
&D_{nm}^{\nu\nu'}=b_{n,n}^{\nu\nu'}\delta_{n,m}+d_{n,m}^{\nu\nu'},\\
\end{eqnarray*}
In the manuscript we use two notations for matrices: letters with a hat, e.g. $\hat{b}$ or square brackets with the element of the matrix inside, e.g.
$[b_{n,m}^{\nu\nu'}]$.

All off-diagonal elements of the matrix $\hat{D}$ are proportional
to the modulation amplitude: $d_{n,m}^{\nu\nu'}\sim g\zeta_{n-m}$ for the corrugated graphene and $d_{n,m}^{\nu\nu'}\sim\alpha_{n-m}$  for the conductivity grating.
The diagonal in diffraction orders matrix $\hat{b}$ is the limit of the matrix $\hat{D}$ if the graphene monolayer is simply flat and homogeneous. It describes the reflection and transmission of the plane wave having the wavevector $(k_n,0,k_{zn})$ by a flat homogeneous monolayer.

In principle, the set of equations (\ref{systF}) can be straightforwardly solved numerically by considering a finite number of spatial field harmonics (see e.g. \cite{bludrev}).
The number of required harmonics should be established for each particular geometry to guarantee the convergency.
This procedure however does not provide neither enough qualitative understanding of the fundamental scattering mechanisms nor yields any simple parametric dependencies of the scattering amplitudes. The analytical treatment allows us to overcome the mentioned limitations of purely numerical analysis.
In the next section we present the analytical solution of the system (\ref{systF}) taking into account the resonance behaviour of the diffracted fields.

\section{Analytical analysis of the diffraction coefficients }

When the wavevectors of one or simultaneously several diffracted waves approximately coincides with the wavevector of the
GPs $k_p$, the determinant of $\hat{D}$ strongly decreases.
In fact, this condition implies that the GPs eigenmodes of the grating are excited. The dispersion relation for these modes is given by $\mathrm{det}\hat{D}=0$. This dispersion relation will be considered in more details in Section 4. Here, in order to separate the resonance and nonresonance diffraction orders, we can first neglect the grating corrections to the dispersion relation and simply set $\mathrm{det}\hat{b}=0$. More explicitly (see details in Appendix A),
\begin{equation}\label{g13}
\mathrm{det} [b_{r,r}^{\nu\nu'}] =\frac{\varepsilon^{(+)}g}{k_{zr}^{(+)}}+\frac{\varepsilon^{(-)}g}{k_{zr}^{(-)}}+2\alpha_0 = 0.
\end{equation}
This is the GP dispersion relation for a homogeneous flat monolayer. For example, in case of a symmetric surrounding $\varepsilon^{(1)}=\varepsilon^{(+)}=1$, the GP wavevector reads $k_p=g\sqrt{1-1/\alpha^2}$. Those diffraction orders for which the condition (\ref{g13}) is approximately fulfilled (``the resonance orders'') are labeled
$r$, $r'$, $r''$, etc. The rest of the diffraction orders (``the nonresonance orders'') are called $N$, $N'$, $N''$, etc.

Accordingly, the matrix $\hat D$ can be decomposed into the four
submatrices: two of them contain the resonance $ \hat R = \left[ {D_{rr'}^{\nu \nu'} } \right]$ and nonresonance $\hat M = \left[ {D_{NN'}^{\nu \nu '} }\right]$
diffraction orders; and the other two a re coupling submatrices, $\hat U = \left[{D_{rN}^{\nu \nu '} } \right]$, $\hat L = \left[{D_{Nr}^{\nu \nu '} } \right]$. We denote the resonant and nonresonant
right-hand sides as $\hat u = \left[ {V_r^\nu  } \right]$ and $\hat
m = \left[ {V_N^\nu  } \right]$ respectively. Decomposing
the submatrix $\hat M$ into a block diagonal, and a
nondiagonal matrix, we have
$$
\hat{M} = \hat{A}(\hat{1} - \hat{\eta})  , \;
A^{\nu\nu'}_{NN'}=\delta_{N,N'}D_{NN}^{\nu\nu'},
$$
where the norm of the matrix $\hat \eta = \hat{A}^{-1}\left[ {d_{NN'}^{\nu\nu'} } \right]$ is small as its elements are
proportional to the small modulation amplitude. Then the inverse to $\hat M$ matrix can be presented in the form of the
series expansion in $\hat \eta$, namely $\hat M^{ - 1}  = \sum\limits_{s =
0}^\infty  {\hat \eta ^s \hat A^{ - 1} } $. As a result, we can solve the nonresonance subsystem for $r_N^\nu$ explicitly
\begin{equation}\label{g18}
\left[ {r_N^\nu  } \right] = \hat M^{ - 1} \left( {\hat m -
\hat L\left[ {r_r^\nu  } \right]} \right).
\end{equation}
Then, substituting the nonresonance amplitude into the resonance subsystem we arrive at the finite system of equations for $r_{r'}^{\nu '}$:
\begin{equation}\label{g19}
\sum\limits_{r',\nu '} {\tilde D_{rr'}^{\nu \nu '}
r_{r'}^{\nu '} }  = \tilde V_r^{\nu},
\end{equation}
where $\left[ {\tilde D_{rr'}^{\nu \nu '} } \right] = \hat
R - \hat U\hat M^{ - 1} \hat L$;  $\left[ {\tilde V_r^\nu  }
\right] = \hat u - \hat U\hat M^{ - 1} \hat m$.
The precision of the solution is defined by the number of retained terms in the matrix  $\hat M^{ - 1}$. For further analytical treatment we will retain terms linear in modulation amplitude in $\tilde V_r^{\nu}$  and quadratic terms in $\tilde D_{rr'}^{\nu \nu '}$. In this approximation it is sufficient to retain the zeroth order term in the series expansion, $\hat M^{ - 1}  \simeq \hat A^{ - 1}$. The explicit expressions for the matrix elements are presented in Appendix A.

In the next section we illustrate the above perturbational method just derived with a simple example.

\section{Illustrative example: first-order resonance under the normal incidence}

A resonant situation of practical interest can take place for normal incidence. Due to the symmetry, two diffraction orders can simultaneously become resonant. Let us consider here the resonance in the first diffraction order, $r=1$, $r'=-1$ on the free-standing ($\varepsilon^{(-)}=\varepsilon^{(+)}=1$) harmonic grating. The resonance for this situation occurs when the Bragg vector of the grating is approximately equal to the real part of the GP wavevector:
\begin{equation}\label{res1}
G \simeq \mathrm{Re}(k_p) = g\mathrm{Re}\left(\sqrt{1-1/\alpha_0^2}\right).
\end{equation}
In what follows we assume the absorption to be small, $\alpha'_0\ll \alpha''_0$, and the period of reciprocal lattice to be large, $G\gg g$. Here and hereafter prime and double prime is used for the real and imaginary part of complex value. This notations should not be mixed with the primes for the integers where they are exclusively used to label diffraction orders.

\subsection{Conductivity modulation of the graphene monolayer}

\begin{figure}
\begin{center}
\includegraphics[width=14cm]{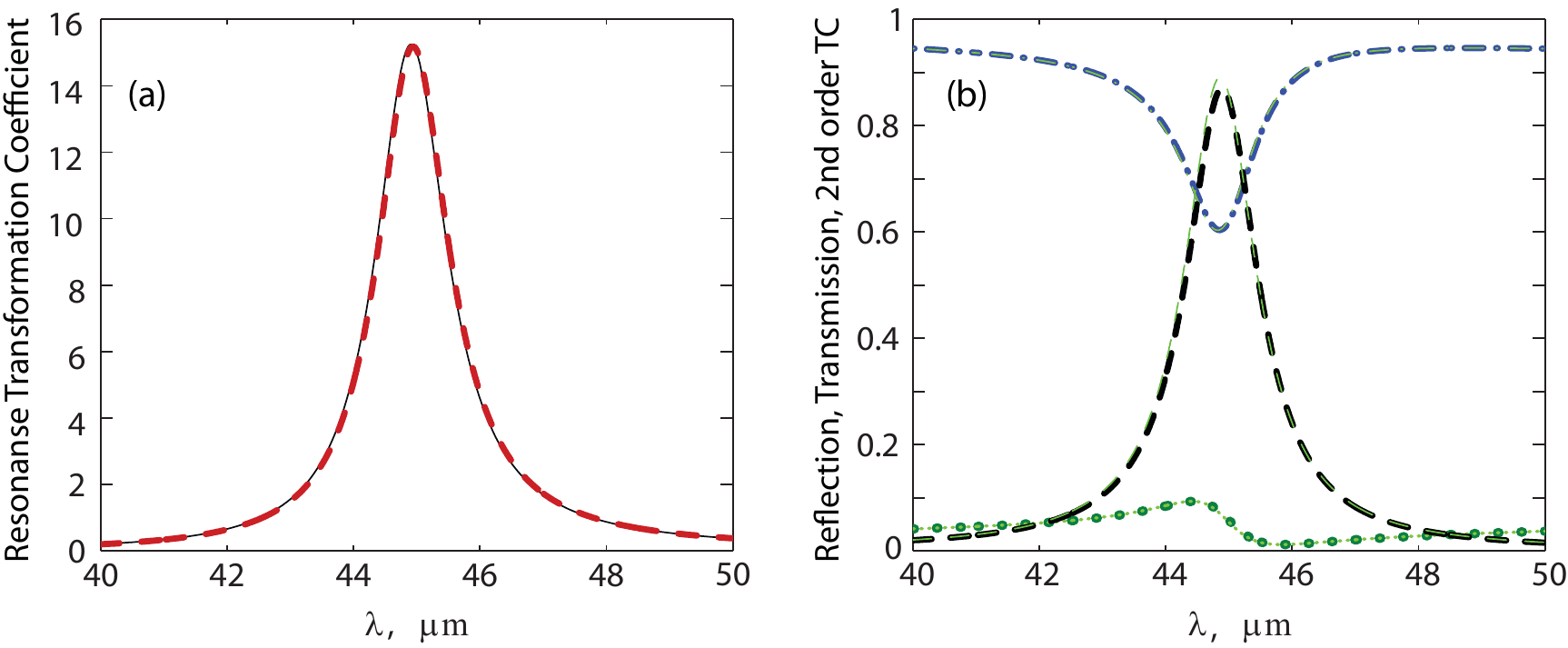}
\caption{(Color online) Spectra of the transformation coefficient for conductivity grating with the modulation amplitude $\mathrm{w}=0.25$. (a) resonant transformation coefficient $|r_{\pm1}^{\nu}|^2$ (thick dashed curve) and (b) reflectance, transmission, and the second order transformation coefficient (dotted, dashed-dotted and dashed thick curves, respectively). Other parameters are: grating period $L=9 \mu$m, chemical potential $\mu=0.4$ eV, relaxation time $\tau=1$ ps.
The calculations performed by means of solution of the eq. set
~(\ref{systF}) are rendered by the thick curves, while the thin curves correspond to the simple analytical expressions (\ref{g37b}) -- (\ref{g38a}).
}\label{graph2-0}
\end{center}
\end{figure}

We take the following spatial dependency of the conductivity
\begin{equation}\label{g35}
\alpha (x) = \alpha_0\left(1 + \mathrm{w}\cos(Gx)\right),
\end{equation}
where, according to Eq.~(\ref{eq1}), $\mathrm{w}=2\alpha_{\pm 1}/\alpha_0$. From Eqs.~(\ref{g19}) after some algebraic derivations we explicitly have the resonance transformation coefficients in the following form:
\begin{eqnarray}\label{g37b}
&r_{\pm1}^{\nu} = -\frac{\alpha_0}{1+\alpha_0}\cdot\frac{\mathrm{w}}{\Delta_r},\\
&\Delta_r = 2\left(\frac{g}{k_{zr}}+\alpha_0 \right) - 2\,\Gamma(\lambda, \mathrm{w}),\nonumber\\
&\Gamma(\lambda, \mathrm{w})=\frac{\alpha_0^2\mathrm{w}^2}{2}\left(\frac{1}{\alpha_0+1} + \frac{1}{2\left(\alpha_0-{ig}/{2G}\right)}\right).\nonumber
\end{eqnarray}
Here $\Gamma(\lambda, \mathrm{w})$ is the quadratic-in-modulation term responsible for the scattering of the resonance wave into neighbouring nonresonance ones. These neighbouring waves (in the main approximation) are the propagating one with $N=0$ and two evanescent ones with $N=\pm2$. The nonresonance field amplitudes are given by Eq.~(\ref{g18}), so that in 0th order they read
\begin{eqnarray}\label{g42}
&r_0^{+} = R_{F} +  \frac{\alpha_0^2\mathrm{w}^2}{\left(\alpha_0+1\right)^2\Delta_r}, \nonumber\\
& r_0^{-} = T_{F} + \frac{\alpha_0^2\mathrm{w}^2}{\left(\alpha_0+1\right)^2\Delta_r},
\end{eqnarray}
while for the second order $\pm 2$ the amplitudes are given by
\begin{equation}\label{g38a}
r_{\pm2}^{\nu} = \frac{\alpha_0^2\mathrm{w}^2}{2\Delta_r\left(1+\alpha_0\right)\left(\alpha_0-{ig}/{2G}\right)}.
\end{equation}
The reflectance and transmittance amplitude coefficients (Fresnel coefficients) of the unmodulated graphene appearing in Eq.~(\ref{g42}) are
\begin{equation*}
R_F=-\frac{\alpha_0}{\alpha_0+1}, \; T_F=\frac{1}{\alpha_0+1}.
\end{equation*}
Fig.~\ref{graph2-0} renders the comparison between the analytically calculated amplitudes according to Eqs.(\ref{g37b})-(\ref{g38a}) and numerical solution of the system (\ref{systF}).

We will now make use of these analytical expressions to find the optimal grating amplitude that provides a maximal intensity of the resonance field.
Separating the real and imaginary parts in denominator $\Delta_r$ in Eq.~(\ref{g37b}), the resonant diffraction amplitude $r_{\pm1}^{\nu}$ can be rewritten as
\begin{equation}\label{rr}
r_{\pm1}^{\nu}= \,\frac{({R_F}/{2})\,\mathrm{w}}{\alpha_0^{\prime}-\Gamma^{\prime}(\lambda, \mathrm{w})-i\left({g}/{k_{zr}^{\prime\prime}}-\alpha_0^{\prime\prime}+\Gamma^{\prime\prime}(\lambda, \mathrm{w})\right)},
\end{equation}
where
\begin{eqnarray*}
&\Gamma^{\prime}(\lambda, \mathrm{w}) \simeq -\frac{\alpha''^2_0\mathrm{w}^2}{2}\left(1+\frac{\alpha'_0}{\alpha''^2_0}\right),\nonumber\\
&\Gamma^{\prime\prime}(\lambda, \mathrm{w})\simeq \frac{\mathrm{w}^2\alpha''_0}{2}.
\end{eqnarray*}
Here we have taken into account that $|\alpha_0|\ll1$ and that in the resonance vicinity on the one hand $k_{zr}\simeq iG$ and on the other hand $k_{zr}\simeq-g/\alpha_0$.  This allows us to make the following simplification inside $\Gamma$: $\alpha_0-{ig}/{2G}\simeq{\alpha_0}/{2}$.

\begin{figure}
\begin{center}
\includegraphics[width=12cm]{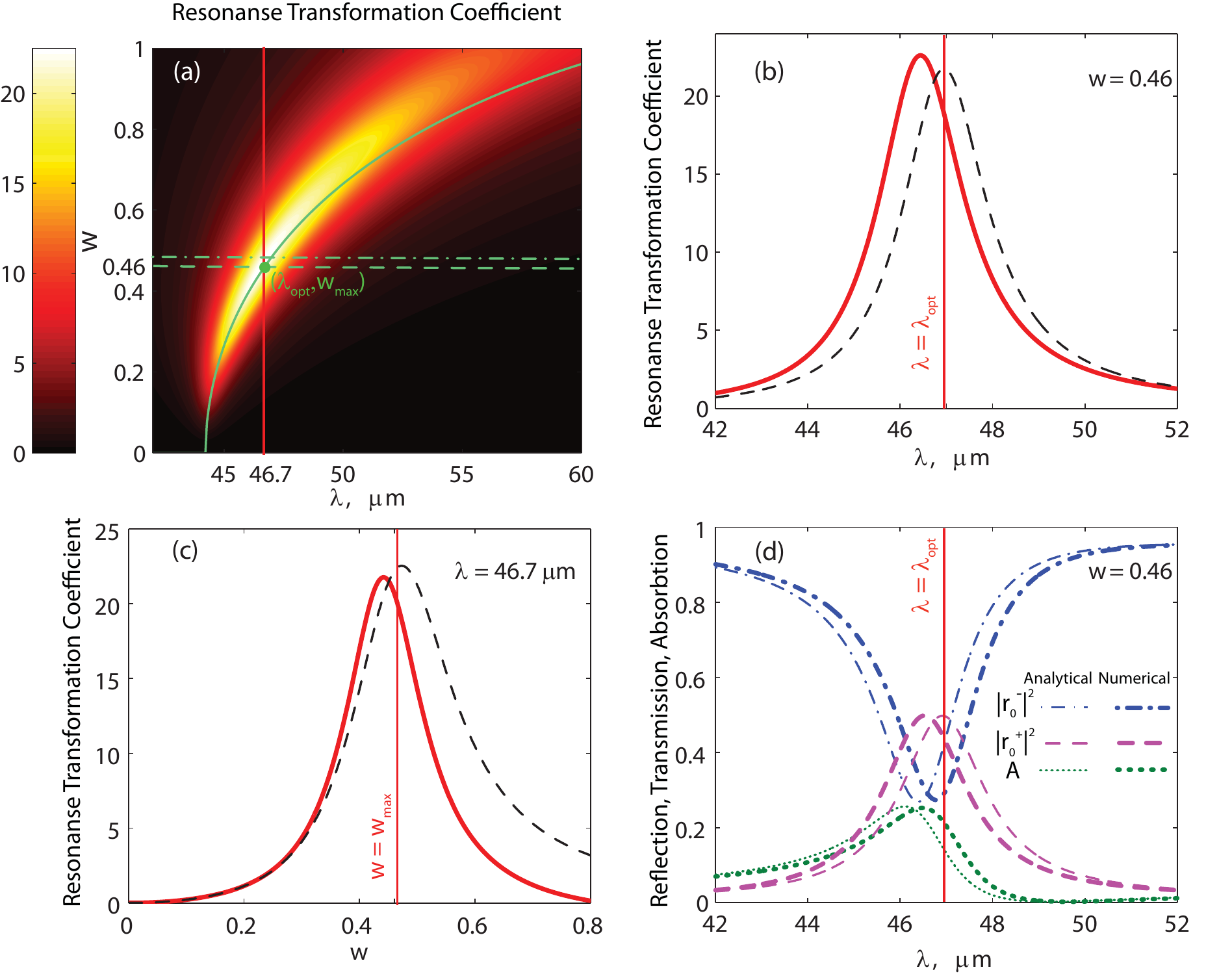}
\caption{(Color online) (a) colorplot showing the resonant transformation coefficient as a function of  the incident wavelength and
modulation amplitude. The continuous, dashed and dash-dotted curves were calculated with the help of (\ref{dispmodif}), (\ref{wopt}) and (\ref{wopt2}) respectively. The intersection of the dashed and continuous curves determines $\lambda$ and $\mathrm{w}$ corresponding to the maximum value of $|r_{\pm1}^{\nu}|^2$ . (b) The wavelength dependence of the resonant transformation coefficient at $\mathrm{w}=\mathrm{w}_{\mathrm{max}}=0.46$; (c) the resonant transformation coefficient as a function of the modulation amplitude at the optimal wavelength
$\lambda=\lambda_{\mathrm{opt}}=46.7\mu m$; in (b,c) the exact calculations and the approximation (\ref{rr}) are rendered by the continuous and discontinuous curves respectively; (d) wavelength spectra of reflectance, transmission, total absorption (shown by dotted, dashed-dotted and discontinuous thick curves respectively) at the optimal modulation amplitude $\mathrm{w}=\mathrm{w}_{\mathrm{max}}=0.46$. The calculations in (d) were performed by means of solution of the eq. set
~(\ref{systF}) (thick curves) and using the simple analytical expressions (\ref{g37b}) -- (\ref{g38a}), thin curves in the plots.
Other parameters are the same as in Fig.~\ref{graph2-0}. Vertical lines in all the panels show the position of the optimal wavelength $\lambda_{\mathrm{opt}}$.}\label{graph2}
\end{center}
\end{figure}

Vanishing the imaginary part of the denominator in Eq.~(\ref{rr}), we get the condition for the maximum of the absolute value of the resonance coefficient:
\begin{equation}\label{dispmodif}
\mathrm{w}_{\mathrm{opt}} - \sqrt{\frac{2}{\alpha_0^{\prime\prime}(\lambda_{\mathrm{opt}})}\left(\alpha_0^{\prime\prime}(\lambda_{\mathrm{opt}})-\frac{g}{k_{zr}^{\prime\prime}(\lambda_{\mathrm{opt}})}\right)}=0,
\end{equation}
so that the expression for $r_{\pm1}^{\nu}$ becomes
\begin{equation}\label{ropt}
r_{\pm1opt}^{\nu}=\frac{R_F}{2} \,\frac{\mathrm{w}_{\mathrm{opt}}}{\alpha_0^{\prime}-\Gamma^{\prime}(\lambda_{\mathrm{opt}}, \mathrm{w}_{\mathrm{opt}})},
\end{equation}
where $\mathrm{w}_{\mathrm{opt}}$ and $\lambda_{\mathrm{opt}}$ in this formula are connected through Eq.~(\ref{dispmodif}). Geometrically, the condition (\ref{dispmodif}) defines a curve in the $(\lambda,\mathrm{w})$
plane. At each point belonging to this curve there are optimum conditions for the resonance GP excitation. At a given wavelength, the optimal grating amplitude can be found from Eq.~(\ref{dispmodif}), and vice versa for a fixed amplitude of the grating Eq.~(\ref{dispmodif}) yields the wavelength. Fig.~\ref{graph2} shows that the curve given by Eq.~(\ref{dispmodif}) (continuous curve in panel (a) of Fig.~\ref{graph2}) indeed goes along the maximum that is seen in the colorplot as a ridge.

Additionally, we can find the value for the modulation amplitude and wavelength at which $|r_{\pm1}^{\nu}|^2$ possesses the global maximum. To do it, we take the derivative of
$|r_{\pm1opt}^{\nu}|^2$ with respect to $\mathrm{w}$ and then equal it to zero. The differentiation can be greatly simplified if instead of $\lambda=\lambda_{\mathrm{opt}}(\mathrm{w}_{\mathrm{opt}})$ given by  Eq.~(\ref{dispmodif}), we substitute $\lambda=\lambda_{0}$, where $\lambda_{0}$ is the resonance wavelength when modulation is neglected. It is given by the approximate condition (\ref{res1}) which can be written as $G\simeq g\mathrm{Re}[i/\alpha_0(\lambda_0)]$ for large $G$. Performing differentiation with the above simplification, $\frac{d}{d\mathrm{w}_{\mathrm{opt}}}|r_{\pm1opt}^{\nu}|^2=0$, we find
\begin{equation}\label{wopt}
\mathrm{w}_{\mathrm{max}}= \frac{\sqrt{2\alpha_0^{\prime}(\lambda_{0})}}{\alpha_0^{\prime\prime}(\lambda_{0})\sqrt{1+{\alpha'_0(\lambda_{0})}/{\alpha''^2_0(\lambda_{0})}}}.
\end{equation}
Assuming the absorption to be small enough, $\alpha'_0\ll\alpha''^2_0$, the expression (\ref{wopt}) can be further simplified:
\begin{equation}\label{wopt2}
\mathrm{w}_{\mathrm{max}}\simeq \frac{\sqrt{2\alpha_0^{\prime}(\lambda_{0})}}{\alpha_0^{\prime\prime}(\lambda_{0})}.
\end{equation}
At this amplitude of the grating the maximum value of $|r_{\pm1opt}^{\nu}|^2$ becomes
\begin{equation*}\label{rrmax}
|r_{\pm1}^{\nu}|_{\rm max}^2\simeq \frac{1}{8\alpha_0^{\prime}\left(1+{\alpha'_0}/{\alpha''^2_0}\right)}\simeq  \frac{1}{8\alpha_0^{\prime}}\gg 1.
\end{equation*}
The squared absolute value of the resonance coefficient is shown in Fig.~\ref{graph2} (b) as function of $\lambda$ at constant $\mathrm{w}$. In contrast, Fig.~\ref{graph2} (c) shows $|r_{\pm1}^{\nu}|^2$ as function of $\mathrm{w}$ at constant $\lambda$. Both approximate and exact numerical calculation are presented. From this comparison we can conclude that, even for moderate modulation amplitudes (for which the optimum is achieved), the analytical approximation is still valid with a reasonable precision and indeed allows predicting the nontrivial conditions for the optimal coupling.

As usual, the resonant excitation of a plasmon is accompanied  by the resonant increase of the absorption.
Taking into account that in our case only one propagating wave is generated by the grating, the absorption is given by $A=1-|r_0^+|^2-|r_0^-|^2$.
The reflection $r_0^-$ and transmission $r_0^+$ amplitude coefficients for the grating amplitude $\mathrm{w}_{\mathrm{max}}$ read approximately
\begin{eqnarray*}
 r_{0max}^+\simeq R_F-\frac{T_F^2}{2}, \,r_{0max}^-\simeq T_F\left(1-\frac{T_F}{2}\right).
\end{eqnarray*}
Then the maximal value of absorption is $A_{\mathrm{\mathrm{max}}}\simeq1/2$. We would like to notice that this is the limiting value of absorption by a monolayer in a symmetric dielectric surrounding \cite{kop2}. The spectra of the transmission, reflection and absorption coefficients are presented in
Fig.~\ref{graph2}. It is seen that the curves plotted with the asymptotic
formulae~ (\ref{g42}) are in a good agreement with ones obtained from the exact solution of the equation set~(\ref{systF}).

\subsection{Relief modulation of the graphene monolayer}

Let us now assume that the interface is corrugated according the following simple law
\begin{equation}\label{1.1b}
z(x)=h\sin(Gx),
\end{equation}
with $h=\pm2i\zeta_{\pm 1}$ being the corrugation amplitude.
In much the same way as in the previous subsection,  simple analytical expressions for field transformation coefficients
can be derived. Using the assumptions $\alpha_0^{\prime}\ll |\alpha_0^{\prime \prime}|$ and $G\gg g$, as before, the resonance transformation coefficients have the following form:
\begin{equation}\label{rrcor}
r_{\pm1}^{\nu}=\frac{igh\alpha_0}{2\Delta_r},\; \Delta_r=\frac{g}{k_{zr}}+\alpha_0+\Gamma(\lambda, h),
\end{equation}
\begin{figure}
\begin{center}
\includegraphics[width=12cm]{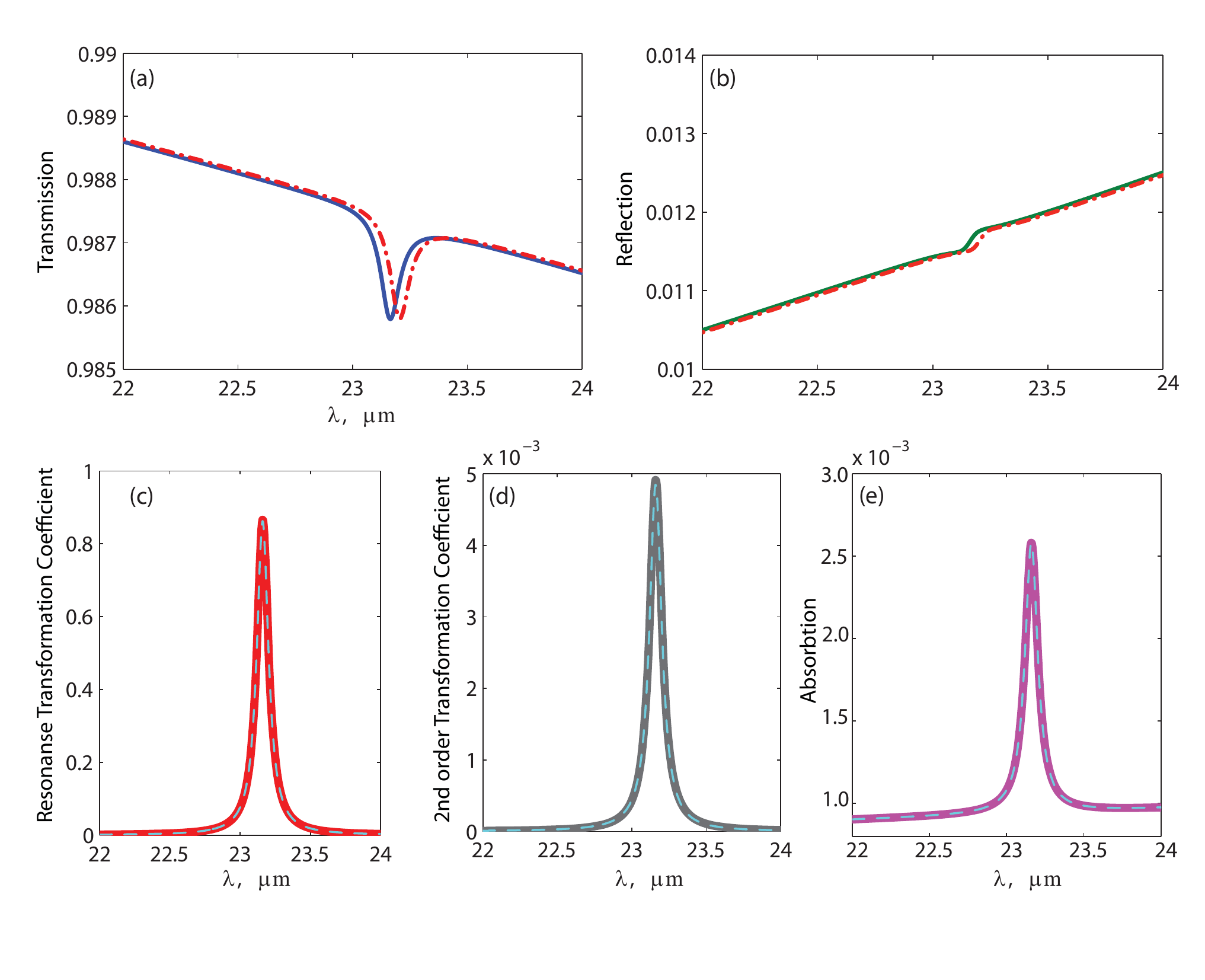}
\caption{(Color online) Dependencies of
(a) reflectivity, (b) transmissivity,  (c) resonant TCs,  (d) TCs in the $\pm 2$-nd
orders and (e) absorption on the incident wavelength. In (a,b) the reflectivity and transmissivity were calculated using both Eq.~(\ref{0cor}) (thick continuous curves) and  first-principle simulations (thin dashed-dotted curves).
 In (c-e) the calculations  were performed according to the equation set~(\ref{systF})(thick continuous curves) and simple analytical expressions (\ref{rrcor}), (\ref{2rcor})(thin discontinuous curves). The parameters taken for textured graphene: the period $L=2.5 \mu$m, grating height $h=30$nm, chemical potential $\mu=0.4$ eV, relaxation time $\tau=6$ ps.}\label{axis}
\end{center}
\end{figure}
where the quadratic term $\Gamma$ in the resonance denominator reads
\begin{equation*}
    \Gamma(\lambda, h)=\frac{(gh)^2}{8}\alpha_0\left(1+2\alpha_0\right).
\end{equation*}
The $0^{\mathrm{th}}$ and $\pm 2^{\mathrm{nd}}$ order transformation coefficients are
\begin{eqnarray}\label{0cor}
&r_0^+=R_F-(gh)^2\frac{\alpha_0^2}{\Delta_r},\\
& r_0^-=T_F+(gh)^2\frac{\alpha_0^2}{\Delta_r}.\nonumber
\end{eqnarray}
\begin{equation}\label{2rcor}
r_{2r}^{\nu}=-\nu\frac{(gh)^2}{\Delta_r}.
\end{equation}
The comparison of the analytical expressions (\ref{rrcor})--(\ref{2rcor}) with numerical simulations by using finite elements method is shown in Fig.~\ref{axis} (a) and (b). We observe a good agreement for sufficiently small amplitudes.
However, unlike the case of conductivity perturbations, the analytical approach for corrugated  gratings with graphene is far more restricted. The approximate solution given by Eq.~(\ref{rrcor})--(\ref{2rcor}) starts to fail when $h$ becomes comparable or larger than 0.25 GP wavelengths. This is also rather different from the case of metallic gratings, in which the perturbational approach is valid for much higher modulation amplitudes (see e.g.~\cite{kats4}). Thus, to fully address the diffraction problem in the region of deeper corrugation gratings, numeric calculations are needed.

\begin{figure}
\begin{center}
\includegraphics[width=12cm]{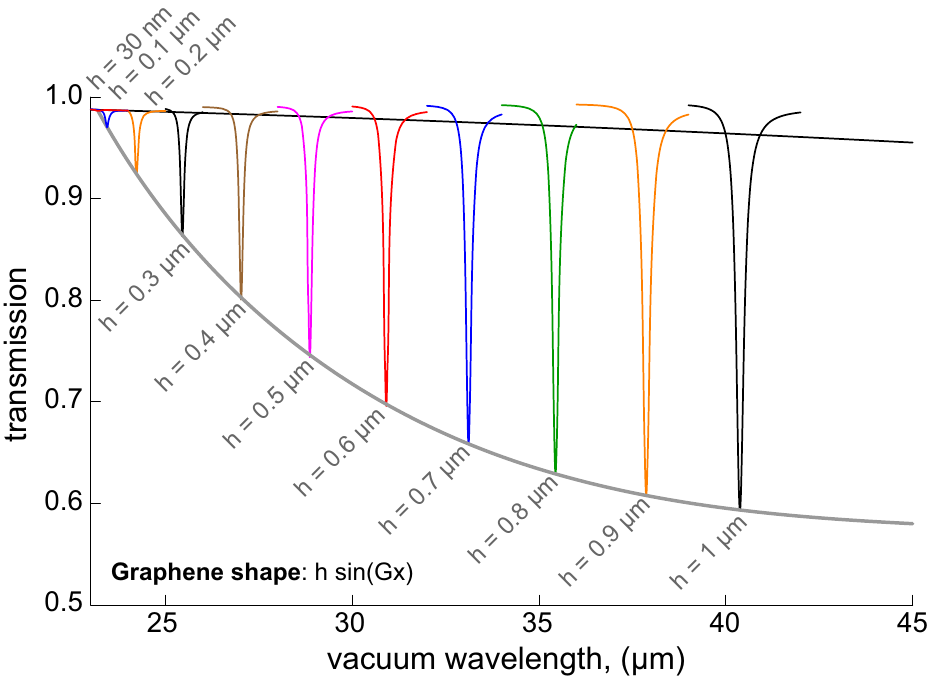}
\caption{(Color online) Transmission spectra for different corrugation amplitudes. The calculations were performed by means of
the first-principle simulations. Black continuous line corresponds to the transmission through the uncorrugated graphene sheet. Gray curve is the envelope for the transmission minima. Parameters for textured graphene:
the period $L=2.5 \mu$m, chemical potential $\mu=0.4$ eV, relaxation time $\tau=6$ ps}\label{href}
\end{center}
\end{figure}

We have performed numeric simulations in a wide range of diffraction grating depths. It is interesting to note that we were unable to find an optimum in dependencies of the scattering coefficients upon $h$. As an example, in Fig.~\ref{href} we show simulated transmission spectra at different depths of the grating. Up to $h$ of order 1$\mu$m (for which $h$ already becomes comparable with GP wavelength) the transmission dip monotonically decreases, without any indication to the existence of the optimal grating amplitude that would provide the best matching between the incident wave and excited GP. This is another peculiarity of the graphene corrugation gratings which is different from both graphene conductivity gratings and metallic relief grating.

\section{Analysis of the GP eigenmodes}

Due to the periodical modulation of graphene, the initially isotropic dispersion relation for graphene plasmons transforms into a bandgap structure.
To find the spectrum of ``grating-dressed'' GP eigenmodes, the homogeneous diffraction problem has to be solved. The resonance perturbation theory developed for the inhomogeneous diffraction problem can be perfectly adapted to the eigenmode solution. Indeed, the resonance diffraction orders have been chosen from the condition of their closeness to the GP eigenmodes. Therefore, vanishing the determinant of the matrix in Eq.~(\ref{g19}) should directly give the perturbed dispersion relation, where both coupling and ``dressing'' (by both inhomogeneous and homogeneous field harmonics) of initially bare GPs are taken into account.

In the excitation problem the mismatch between the wavevector of the incident wave $\mathbf{k}_i=(k,0,k_{z})$ and the plasmon wavevector $\mathbf{k}_p=(k_p,0,k_{zp})$ can be overcome via the reciprocal lattice vector $\mathbf{G}=(G,0,0)$:
\begin{equation}\label{kp1}
\pm k_p=\pm k_r=k+rG,
\end{equation}
where  $r= \pm1, \pm2, \ldots$;  $r>0$ and $r<0$ correspond to the forward and backward propagation of the excited GPs respectively.
In the homogenous problem the plasmon wavector $q$ should be taken instead of one of the resonance wavevectors, so that the Bragg vector provides the coupling between bare GPs and its scattering via different propagating and evanescent diffraction orders. As opposed to $k_p$, being the GP wavevector for the unperturbed graphene, $q$ stays for the GP wavevector for the grating. In our plane geometry only two bare plasmons with the wavevectors, $k_p$ and $-k_p$ can be simultaneously coupled. Then this coupling implies the following approximate condition for their wavevectors:
\begin{equation}\label{kp2}
\mp k_p=\pm k_p+mG.
\end{equation}
Thus, in order to pass form the inhomogeneous problem to the inhomogeneous one, we can change the tangential component
of the wavevector of one of the resonance field harmonics to GP wavevector, e.g. $k_r\rightarrow q$. Then the rest of the wavevectors become $k_{r+n}\rightarrow q +nG$. After this change we can use the equation
\begin{equation}\label{det11}
\mathrm{det}[\tilde{D}_{rr'}^{\nu\nu'}] = 0
\end{equation}
as an approximate dispersion relation for GPs on the grating. This approximation works in the vicinity of the point where the bare GPs dispersion curves intersect, that is in the vicinity of a certain bangap.

The equation (\ref{det11}) is explicitly written in Appendix B. Without entering the mathematical details, let us discuss here the main scattering mechanisms.
The elements of the matrix $[\tilde{D}_{rr'}^{\nu\nu'}]$ contain terms quadratic in modulation amplitude. They mainly contribute to a simultaneous nonlinear (in modulation amplitude) shift of the bare dispersion curves and to the decay rate (nonlinear ``widening''). In contrast, the linear terms basically affect the splitting between the bare dispersion branches, and also increase the GPs decay rate. In what follows we will consider the linear approximation (i.e. we neglect all quadratic-in-modulation amplitude terms in the matrix elements). This is reasonable if the coupling harmonic amplitude $\zeta_{r-r'}$ (or $\alpha_{r-r'}$) obeys the inequality  $|\zeta_r|\gg|\zeta_N|^2$ (or $|\alpha_r|\gg|\alpha_N|^2$) for any $N$.

For simplicity, consider the coupling between initial bare GPs via first-order scattering by a harmonic grating. We take the equivalent double resonance diffraction problem considered in Section 4, where the resonance diffraction orders were $r=1$, $r'=-1$. However, in the homogeneous problem in order to be at the first Brillouin zone edge, it is more appropriate to take $r=0$, $r'=-1$. We set $k_0\rightarrow q$ and $k_{-1}\rightarrow q-G$. Additionally, in order to simplify the equations writing we will assume the symmetric surrounding, $\varepsilon^{(-)}=\varepsilon^{(+)}=1$.

\subsection{Conductivity grating}

We consider the simplest conductivity grating given by Eq.~(\ref{g35}).
In order to estimate the splitting and decay rate at the very edge of the Brillouin zone we set $q=G/2$. Then the dispersion relation reads (see details in Appendix B):
\begin{eqnarray}\label{displaw}
\left(\frac{g}{q_{0z}}+\alpha_0\right)^2-\alpha_{-1}\alpha_{1}=0,
\end{eqnarray}
where $q_{0z}=\sqrt{g^2-G^2/4}$.  We have to assume the frequency to be complex-valued, $\omega=\omega'+i\omega''$ (or $g=g'+ig''$). The solution of this equation has two complex roots, $\omega^+$ and $\omega^-$ which are frequencies of the upper and lower split GP branches
\begin{equation*}\label{bandgapCond2}
\omega^{\pm}=c\frac{\pi}{L}\left(\alpha_0^{\prime\prime}(\omega_0)-i\alpha_0^{\prime}(\omega_0)\right)\left(1\pm\mathrm{w}/2\right),
\end{equation*}
where $\omega_0$ is the frequency of GP in a homogeneous free-standing flat graphene monolayer following from the dispersion relation (\ref{g13}). Extracting then the real and imaginary part of Eq.~(\ref{displaw}), we obtain the splitting $\Delta\omega = \omega'^+-\omega'^-$:
\begin{equation*}\label{bandgapCond2}
\bigtriangleup\omega=c\frac{\pi}{L}\alpha_0^{\prime\prime}(\omega_0)\mathrm{w}
\end{equation*}
and the imaginary frequency components
\begin{equation}\label{bandgapCond1}
\omega''^\pm =-c\frac{\pi}{L}\alpha_0^{\prime}(\omega_0)\left(1\pm\mathrm{w}/2\right).
\end{equation}
Taking the optimal modulation amplitude $\mathrm{w}=\mathrm{w}_{\mathrm{max}}=0.46$, and other parameters as in Fig.~2,  the decay rates corresponding to Eq.~(\ref{bandgapCond1}) are 1.51 ps and 2.42 ps for the upper and lower branches respectively.

\subsection{Relief modulation of the graphene monolayer}

We assume a relief profile to be given by Eq.~(\ref{1.1b}). At the Brillouin zone edge $q=G/2$
the dispersion relation for corrugated graphene in linear approximation is given by (see Appendix B)
\begin{eqnarray}\label{displawRaab}
\left(\frac{g}{q_{0z}}+\alpha_0\right)^2-\alpha_0^2g^2U_{0,-1}U_{-1,0}\zeta_{1}\zeta_{-1}=0,\\
U_{r,r'}=\frac{1}{g}\left(q_{z0}+\frac{G q}{q_{z0}}\right), \; r,r'=0, \,-1.\nonumber
\end{eqnarray}
Taking into account that on the one hand $q_{0z}\simeq iG$ and on the other hand $q_{0z}\simeq-g/\alpha_0$,  we obtain the following expression for the coefficients $U_{-1,0}$ and $U_{0,-1}$:
\begin{eqnarray*}
U_{-1,0}=U_{0,-1}\simeq -\frac{i}{\alpha_0^{\prime\prime}}\left(1+\alpha_0^{\prime\prime}\frac{G}{g}\right).\nonumber
\end{eqnarray*}
This allows us to simplify the dispersion relation~(\ref{displawRaab}):
\begin{eqnarray}\label{displawR}
\left(\frac{g}{q_{0z}}+\alpha_0\right)^2-\alpha_0^2g^2\frac{\zeta_{1}\zeta_{-1}}{\alpha_0^{\prime\prime 2}}\left(1+\alpha_0^{\prime\prime}\frac{G}{g}\right)^2=0.
\end{eqnarray}
Following the same procedure as in previous subsection, we obtain the frequencies of the upper and lower split GP branches
\begin{eqnarray}\label{bandgapRel2}
&\omega^{\pm}=c\frac{\pi}{L}\left(\alpha_0^{\prime\prime}(\omega_0)-i\alpha_0^{\prime}(\omega_0)\right)\times\\
&\left(1\pm\left(1+\frac{G}{g_0}\alpha_0^{\prime\prime}(\omega_0)\right)\frac{g_0h}{2\alpha_0^{\prime\prime}(\omega_0)}\right),\nonumber
\end{eqnarray}
from which we find the value of the band gap
\begin{equation*}
\bigtriangleup\omega
=c\frac{\pi}{L}\left(1+\frac{G}{g_0}\alpha_0^{\prime\prime}(\omega_0)\right)\frac{g_0h}{\alpha_0^{\prime\prime}(\omega_0)}\nonumber,
\end{equation*}
with $g_0=\omega_0/c$.
The imaginary part of the equation (\ref{bandgapRel2}) provides the the imaginary frequency components
\begin{eqnarray}\label{bandgapCond12}
\omega''^\pm =-c\frac{\pi}{L}\alpha_0^{\prime}(\omega_0)\times\nonumber\\
\left(1\pm\left(1+\frac{G}{g_0}\alpha_0^{\prime\prime}(\omega_0)\right)\frac{g_0h}{2\alpha_0^{\prime\prime}(\omega_0)}\right),
\end{eqnarray}
For the chemical potential, the relaxation time and the period of the lattice  as in Fig.~\ref{axis}, the decay rates for the upper and lower branches are 5.574 ps and 5.584 ps respectively.

\section{Conclusions}

In this paper we have used resonant perturbation theory to analytically solve the diffraction problem for graphene gratings. Both interface corrugation and periodical change in optical conductivity have been considered. We have provided simple analytical expressions for the amplitudes of scattered plane waves in different diffraction orders. For the case of graphene with  modulated optical conductivity we have found the optimal modulation amplitude and wavelength corresponding to the best matching of the incident wave and graphene plasmons. On the other hand, we have shown that the surface relief grating does not have an optimal depth. Instead, it shows a monotonous increase of its efficiency with the increase of the depth. We have also studied the dispersion relation of graphen plasmons on the grating. We have found the value of the band-gap at the edge of the first Brilluoin zone and the decay rates of the split GP modes.

We have considered the simplest case of one-dimensional harmonic periodicity for illustrative purposes,
but the approach allows a generalization to two-dimensional and multilayered graphene periodic structures.

\appendix
\setcounter{section}{1}
\section*{Appendix A}

The vectorial components of the electromagnetic fields in the superstrate/substrate can be written in the Fourier-Floquet expansion form
 \begin{eqnarray}\label{field11}
 &\left[
\begin{array}{c}
   E_{x}^{\nu}(x,z) \\
   E_{z}^{\nu}(x,z) \\
   H^{\nu}(x,z)\\
 \end{array}
 \right] =
  \delta_{\nu,+}\left[
\begin{array}{c}
   1 \\
   \displaystyle\varepsilon^{(+)}\frac{k}{k_{z}}\\
    -\displaystyle\varepsilon^{(+)}\frac{g}{k_{z}}\\
 \end{array}
 \right] e^{i(kx - k_{z}z)}+\nonumber\\
&+\sum_n
 \left[
 \begin{array}{c}
     1 \\
   -\displaystyle\varepsilon^{(\nu)}\frac{\nu  k_n}{k_{zn}^{(\nu)}}\\
   \displaystyle\varepsilon^{(\nu)}\frac{\nu g}{k_{zn}^{(\nu)}}
 \end{array}
 \right]r_{n}^{\nu}e^{i(k_nx+\nu k_{zn}^{(\nu)}z)} ,
\end{eqnarray}
where $r_{n}^{\nu}$ are the transformation coefficients, the subscript $\nu$ stays for the field in the superstate, $\nu=+$, or in the substrate, $\nu=-$.

Matching the electromagnetic fields at the interface, we get the equation set (\ref{systF}) for transformation coefficients. The matrix $\hat{D} \equiv [ D_{nm}^{\nu\nu'}]$ presents the sum of the diagonal and non-diagonal matrices:
$
    D_{nm}^{\nu\nu'}=b_{n,n}^{\nu\nu'}\delta_{n,m}+d_{n,m}^{\nu\nu'}.
$
The diagonal matrix $\hat{b}=[b_{n,n}^{\nu\nu'}]$ corresponds to the set of equations for the diffraction at the homogeneous flat graphene sheet:
\begin{eqnarray}\label{DCond}
    &b_{n,n}^{\nu\nu'}=\left[
                         \begin{array}{cc}
                          1 & -1 \\
                         \frac{\varepsilon^{(+)}g}{k_{zn}^{(+)}} & \left(\frac{\varepsilon^{(-)}g}{k_{zn}^{(-)}}+2\alpha_0\right)\\
                         \end{array}
                     \right].
\end{eqnarray}
The non-diagonal matrix and the right-hand side of the equation set (\ref{systF}) for the case of conductivity grating have the following form
    \begin{eqnarray}\label{DCond2}
    &d_{n,m}^{\nu\nu'}=\left[
                         \begin{array}{cc}
                          0 & 0 \\
                          0 & 2\alpha_{n-m}\\
                         \end{array}
                     \right], \nonumber\\
    &V_n^{\nu}=\left[
                  \begin{array}{c}
                    -1 \\
                    \frac{\varepsilon^{(+)}g}{k_{zn}^{(+)}} \\
                  \end{array}
                \right]\delta_{n,0};
    \end{eqnarray}
while for the relief grating they read
\begin{eqnarray}\label{DRel1}
    &d_{n,m}^{\nu\nu'}=\left[
                         \begin{array}{cc}
                           U_{n,m}^{(+)} & U_{n,m}^{(-)} \\
                          \varepsilon^{(+)} & -\left(\varepsilon^{(-)}+2\alpha_0U_{n,m}^{(-)}\right) \\
                         \end{array}
                     \right]g\zeta_{n-m},\nonumber\\
   &V_n^{\nu}=\left[
                  \begin{array}{c}
                    -1 \\
                     \frac{{\varepsilon^{(+)}g}}{{k_{zn}^{(+)}}} \\
                  \end{array}
                \right]\delta_{n,0}+\left[
                  \begin{array}{c}
                    U_{n,0}^{(+)}\\
                    -\varepsilon^{(+)}\\
                  \end{array}
                \right]g\zeta_{n},
\end{eqnarray}
where
\begin{eqnarray*}
    U_{n,m}^{(\pm)}=\frac{1}{g}\left(k_{zm}^{(\pm)}-\frac{G (n-m) k_m}{k_{zm}^{(\pm)}}\right).\nonumber
\end{eqnarray*}

To solve the infinite set of equations (\ref{systF}) for
$r_n^{\nu}$, we use resonance perturbation theory described in the Section~\ref{2}. In the main approximation (keeping only the zero-order term in the series expansion)
matrix $\hat{M}^{-1}$ reads:
\begin{eqnarray}\label{MM}
\hat{M}^{-1} \simeq\frac{1}{\Delta_N}\left[
  \begin{array}{cc}
   \frac{\varepsilon^{(-)}g}{k_{zN}}+2\alpha_0 & 1\\
    -\frac{\varepsilon^{(+)}g}{k_{zN}} & 1\\
  \end{array}
\right],
\end{eqnarray}
where $\Delta_N=\varepsilon^{(+)}g/k_{zN}^{(+)}+\varepsilon^{(-)}g/k_{zN}^{(-)}+2\alpha_0$.
Thus, the renormalized matrix $[\tilde{D}_{rr'}^{\nu\nu'}]$ can be written as
$
\tilde{D}_{rr'}^{\nu\nu'}=D_{rr'}^{\nu\nu'}-\gamma_{r,r'}^{\nu\nu'}
$,
where $\gamma_{r,r'}^{\nu\nu'}$ for the conductivity grating is given by the expression:
\begin{eqnarray}\label{gamrel}
    \gamma_{r,r'}^{\nu\nu'}= \left[\begin{array}{cc}
                                                     0 & 0 \\
                                                     0 & 4\sum\limits_N\frac{\alpha_{r-N}\alpha_{N-r'}}{\Delta_N} \\
                                                   \end{array}\right].
\end{eqnarray}
and for the relief grating it is somewhat more cumbersome:
\begin{eqnarray}\label{gamcon}
  &\gamma_{r,r'}^{++}=g^2\sum\limits_N\frac{\zeta_{r-N}\zeta_{N-r'}}{\Delta_N} \Biggl\{\varepsilon^{(+)}\left(U_{r,N}^{(+)}+U_{r,N}^{(-)}\right)+ \nonumber\\
  &+ U_{N, r'}^{(+)}\left[U_{r,N}^{(+)}\left(2\alpha_0+\frac{\varepsilon^{(-)}g}{k_{zN}^{(-)}}\right)+\frac{\varepsilon^{(+)}g}{k_{zN}^{(+)}}U_{r,N}^{(-)}\right]\Biggr\},\nonumber\\
  &\gamma_{r,r'}^{+-}=g^2\sum\limits_N\frac{\zeta_{r-N}\zeta_{N-r'}}{\Delta_N}\Biggl\{-\left(\varepsilon^{(-)}+2\alpha U_{N, r'}^{(-)}\right)\times\nonumber\\
  &\left(U_{r,N}^{(+)}+U_{r,N}^{(-)}\right)+\\
  &+U_{N,r'}^{(-)}\left[U_{r,N}^{(+)}\left(2\alpha_0+\frac{\varepsilon^{(-)}g}{k_{zN}^{(-)}}\right)+\frac{\varepsilon^{(+)}g}{k_{zN}^{(+)}}U_{r,N}^{(-)}\right]\Biggr\},\nonumber\\
  &\gamma_{r,r'}^{+-}=g^2\sum\limits_N\frac{\zeta_{r-N}\zeta_{N-r'}}{\Delta_N}\Biggl\{\varepsilon^{(+)}\left(\varepsilon^{(+)}-\varepsilon^{(-)}-2\alpha U_{r,N}^{(+)}\right)+\nonumber\\
  &+\varepsilon^{(+)}U_{N,r'}^{(+)}\left[\frac{\varepsilon^{(-)}g}{k_{zN}^{(-)}}+2\alpha_0+\frac{g}{k_{zN}^{(+)}}\left(\varepsilon^{(-)}+2\alpha U_{r,N}^{(+)}\right)\right]\Biggr\},\nonumber\\
  &\gamma_{r,r'}^{--}=g^2\sum\limits_N\frac{\zeta_{r-N}\zeta_{N-r'}}{\Delta_N}\Biggl\{-\left(\varepsilon^{(-)}+2\alpha U_{N,r'}^{(+)}\right)\times\nonumber\\
  &\left(\varepsilon^{(+)}-\varepsilon^{(-)}-2\alpha U_{r,N}^{(+)}\right)+\nonumber\\
  &+\varepsilon^{(+)}U_{N,r'}^{(-)}\left[\frac{\varepsilon^{(-)}g}{k_{zN}^{(-)}}+2\alpha_0+\frac{g}{k_{zN}^{(+)}}\left(\varepsilon^{(-)}+2\alpha U_{r,N}^{(+)}\right)\right]\Biggr\}.\nonumber
     \end{eqnarray}
The right-hand side of (\ref{g19}) in the main approximation is
\begin{eqnarray}\label{systF2}
   \tilde{V}_r^{\nu}=\left[
                         \begin{array}{c}
                           0 \\
                          -4\alpha_r \frac{\varepsilon^{(+)}g}{k_{z}\Delta_0}\\
                         \end{array}
                       \right]
\end{eqnarray}
for the conductivity grating and
\begin{eqnarray}\label{systF1}
\tilde{V}_r^{\nu}=g\zeta_r\left[
 \begin{array}{c}
  U_{r,0}^{(+)}+\frac{1}{\Delta_0}\left(U_{r,0}^{(+)}\left(\frac{\varepsilon^{(-)}g}{k_{z}}-\frac{\varepsilon^{(+)}g}{k_{z}}+2\alpha\right)-2\frac{\varepsilon^{(+)}g}{k_{z}}U_{r,0}^{(-)}\right) \\
   \frac{1}{\Delta_0}\left(4\alpha U_{r,0}^{(+)}-\varepsilon^{(-)}+\varepsilon^{(+)}\right)\\
   \end{array}
   \right]
\end{eqnarray}
for the interface corrugation.

In this paper we only consider the normal incidence onto a harmonically modulated graphene sheet~(\ref{g35}), (\ref{1.1b}) in the vacuum surrounding $\varepsilon^{(-)}=\varepsilon^{(+)}=1$. Under the assumption of a large Bragg vector, $G\gg g$, the tangential and normal components of the wavevectors of the diffracted waves in the $\pm1^{\mathrm{st}}$, $\pm2^{\mathrm{nd}}$- and $0^{\mathrm{th}}$ orders read explicitly
\begin{eqnarray}\label{kxkz}
& k = k_0 = 0, \; k_{r} = \pm G, \; k_{\pm2} = \pm 2G,\\
& k_{z} = k_{z0} = 1, \; k_{zr} =\pm  i G , \; k_{z\pm2} =\pm  2iG,\nonumber
\end{eqnarray}
Assuming also $k_{zr}\simeq-g/\alpha_0$, the coefficients $U_{n,m}$ can be simplified:
\begin{eqnarray}\label{uu}
  &U_{r,0}=1, \;  U_{1,2}=U_{-1,-2}=i G=-\frac{1}{\alpha_0}, \nonumber\\
  &U_{0,r}=0, \; U_{2,1}=U_{-2,-1}=0,
\end{eqnarray}
Substituting expressions (\ref{kxkz}) and (\ref{uu}) in Eqs.~(\ref{gamrel}) -- (\ref{systF1}), we obtain the following expression for the renormalized matrix, $[\tilde{D}_{rr'}^{\nu\nu'}]$, and the right-hand side $[\tilde{V}_{r}^{\nu}]$, for the relief grating:
\begin{eqnarray}\label{systF11}
&\tilde{D}_{rr}^{\nu\nu'}=D_{rr'}^{\nu\nu'}-(gh)^2\left[\begin{array}{cc}
                                                             \left(1-\alpha_0\right) & -\left(1-\alpha_0\right)  \\
                                                             -2\alpha_0 & -2\alpha_0^2 \\
                                                           \end{array}
                                                         \right],\nonumber\\
&\tilde{V}_r^{\nu}=\frac{g\zeta_r}{1+\alpha_0}\left[
                         \begin{array}{c}
                           2\alpha_0\\
                           2\alpha_0\\
                         \end{array}
                       \right]
\end{eqnarray}
and for the conductivity grating:
\begin{eqnarray}\label{systF21}
    &\tilde{D}_{rr'}^{\nu\nu'}=D_{rr'}^{\nu\nu'}-\left[ \begin{array}{cc}
                                                                            0 & 0 \\
                                                                            0 & \left(\frac{\alpha_0^2\mathrm{w}^2}{\alpha_0+1} + \frac{\alpha_0^2\mathrm{w}^2}{2\left(\alpha_0-{i}/{2G}\right) }\right)\\
                                                                          \end{array}
                                                                        \right], \nonumber\\
    &\tilde{V}_r^{\nu}=\left[\begin{array}{c}
     0 \\
      -\frac{\mathrm{w}\alpha_0}{1+\alpha_0}\\
      \end{array}\right]
\end{eqnarray}
It is also worth noticing that for a strictly normal incidence the resonance matrix possesses the following symmetry property: $\tilde{D}_{rr}^{\nu\nu'}=\tilde{D}_{-r-r}^{\nu\nu'}$. This
property results in the $ r_r^{\nu}=r_{-r}^{\nu}$ which substantially simplify the solution of the resonance subsystem. Substituting Eqs. (\ref{systF11}) and (\ref{systF21}) into Eqs.
(\ref{g18}) and (\ref{g19}) we derive the resonance, $r_r^{\nu}$, and nonresonance,  $r_N^{\nu}$, transformation coefficients for the case of conductivity and interface modulation.

\appendix
\setcounter{section}{2}
\section*{Appendix B}

Taking into account the quadratic-in-modulation amplitude terms $\gamma$,
the dispersion relation (\ref{det11}) (for the diffraction orders $r, r'=0,\, -1$)  ca be explicitly written as
\begin{eqnarray}\label{det2}
   \hspace{-3.5 cm} \mathrm{det}\left[
  \begin{array}{cccc}
    1+\gamma_{0,0}^{++} &   -1+\gamma_{0,0}^{+-}  & d_{0,-1}^{++} +\gamma_{0,-1}^{++}  & d_{0,-1}^{+-} +\gamma_{0,-1}^{+-} \\
    \varepsilon^{(+)}g/q_{0z}^{(+)}+\gamma_{0,0}^{-+} &  \varepsilon^{(-)}g/q_{0z}^{(-)}+2\alpha+\gamma_{0,0}^{--} &  d_{0,-1}^{-+} +\gamma_{0,-1}^{-+} & d_{0,-1}^{--} +\gamma_{0,-1}^{--} \\
    d_{-1,0}^{++} +\gamma_{-1,0}^{++}  & d_{-1,0}^{+-} +\gamma_{-1,0}^{+-}  &  1+\gamma_{-1,-1}^{++} & -1+\gamma_{-1,-1}^{+-}  \\
    d_{-1,0}^{-+} +\gamma_{-1,0}^{-+} &  d_{-1,0}^{--} +\gamma_{-1,0}^{--} &  \varepsilon^{(+)}g/q_{-1z}^{(+)}+\gamma_{-1,-1}^{-+}  & \varepsilon^{(-)}g/q_{-1z}^{(-)}+2\alpha+\gamma_{-1,-1}^{--}  \\
  \end{array}
\right]=0,\nonumber\\
\end{eqnarray}
where $q_{0z}^{(\pm)}=\sqrt{\varepsilon^{(\pm)}g^2-q^2}$ and $q_{-1z}^{(\pm)}=\sqrt{\varepsilon^{(\pm)}g^2-(q-G/2)^2}$; $d_{rr'}^{\nu\nu'}$ and $\gamma_{rr'}^{\nu\nu'}$ for corrugation grating are given by (\ref{DRel1}) and (\ref{gamcon}), while for the conductivity grating by (\ref{DCond2}) and (\ref{gamrel}).

In the linear approximation (i.e. neglecting quadratic-in-modulation amplitude terms, $\gamma$), Eq.~(\ref{det2}) strongly simplifies. For the conductivity grating it transforms to
\begin{eqnarray}\label{displaw11}
\left(\frac{g}{q_{0z}}+\alpha_0\right)\left(\frac{g}{q_{-1z}}+\alpha_0\right)-\alpha_{-1}\alpha_{1}=0
\end{eqnarray}
while for the corrugation grating it becomes
\begin{eqnarray}\label{displawR11}
&\left(\frac{g}{q_{0z}}+\alpha_0\right)\left(\frac{g}{q_{-1z}}+\alpha_0\right)-\nonumber\\
&-\alpha_0^2g^2U_{0,-1}U_{-1,0}\zeta_{1}\zeta_{-1}=0.
\end{eqnarray}
In order to estimate the splitting at the very edge of the Brillouin zone we have to set $q=G/2$, which implies $q_{0z}=q_{-1z}$ in Eqs.~(\ref{displaw11}),~(\ref{displawR11}).

\end{document}